\documentclass{article}
\usepackage{spconf,amsmath,graphicx}

\usepackage{amssymb,amsmath,epsfig}
\usepackage{textcomp}
\usepackage{amsfonts,amsthm,amsopn}
\usepackage{hyperref}
\usepackage{adjustbox}
\usepackage{tikz}
\newcommand*\circled[1]{\tikz[baseline=(char.base)]{
    \node[shape=circle, draw, inner sep=0.5pt,
minimum height=10pt] (char) {\vphantom{1g}#1};}}

\usepackage{url}
\usepackage[capitalise]{cleveref}
\usepackage{graphicx,caption,lipsum}
\usepackage{booktabs,multirow}
\usepackage{placeins}
\usepackage{algpseudocode,algorithm}
\usepackage{color}
\usepackage{color}
\usepackage{enumerate}   
\usepackage[normalem]{ulem}

\ninept
\def\reg{{\rm\ooalign{\hfil
      \raise.07ex\hbox{\scriptsize R}\hfil\crcr\mathhexbox20D}}}

\title{Disentangled Training with Adversarial Examples For Robust Small-footprint Keyword Spotting}

\makeatletter
\def\name#1{\gdef\@name{#1\\}}
\makeatother 
\name{Zhenyu Wang$^{1*}$ \thanks{$^*$Work performed while the author was an intern at META.}, Li Wan$^2$, Biqiao Zhang$^2$, Yiteng Huang$^2$, Shang-Wen Li$^2$, \\ Ming Sun$^2$, Xin Lei$^2$, Zhaojun Yang$^2$}

\address{
 $^1$ The University of Texas at Dallas, USA\\
 $^2$ META AI \\
{\small \tt}}

\newcommand{\vct}[1]{\boldsymbol{\mathbf{#1}}} 

\begin{document}

\maketitle

\begin{abstract}
A keyword spotting (KWS) engine that is continuously running on device is exposed to various speech signals that are usually unseen before. It is a challenging problem to build a small-footprint and high-performing KWS model with robustness under different acoustic environments. In this paper, we explore how to effectively apply adversarial examples to improve KWS robustness. We propose datasource-aware disentangled learning with adversarial examples to reduce the mismatch between the original and adversarial data as well as the mismatch across original training datasources. The KWS model architecture is based on depth-wise separable convolution and a simple attention module. Experimental results demonstrate that the proposed learning strategy improves false reject rate by $40.31\%$ at $1\%$ false accept rate on the internal dataset, compared to the strongest baseline without using adversarial examples. Our best-performing system achieves $98.06\%$ accuracy on the Google Speech Commands V1 dataset.
\end{abstract}


\section{Introduction}
\label{sec:intro}

With the proliferation of voice assistant devices, the development of efficient and accurate keyword spotting (KWS) systems has attracted much attention in the literature. These devices rely heavily on an on-device KWS which correctly `triggers' the system to send audio into the cloud for interpretation. Due to constrained hardware resources, on-device KWS systems need to achieve high performance in various acoustic environments with a small memory footprint and high computation efficiency. Advances in this area are a considerable influence on the perceived user experience of the device, as a failure to wake up on a trigger attempt is frustrating and a false wake on ambient noise can be considered as an infraction of user privacy.

Conventional approaches to KWS are based on large vocabulary continuous speech recognition (LVCSR), targeting efficient keyword search from the lattices \cite{miller2007rapid,103088}. However, the LVCSR-based technique often consumes high computational resources and is inapplicable to on-device KWS systems. Hidden Markov Models (HMM) have been a commonly used technique for building small-footprint and low-latency KWS systems, which takes either keyword \cite{rohlicek1989continuous,rose1990hidden} or monophone \cite{wu2018monophone,sun2017compressed} as modeling units. More recently, deep neural networks (DNNs) have been adopted to predict word units in the keyword for each frame \cite{chen2014small}.
Convolutional neural networks (CNNs) have seen great success in small-footprint tasks due to its effective representation of time-frequency structure and low memory bandwidth requirements \cite{sainath2015convolutional} \cite{tang2018deep} \cite{coucke2019efficient}.

A KWS system is continuously exposed to various audio signals that are mostly unseen before. Due to the high variability of the acoustic environment in the aspects of SNRs, noise types, and accents, it is a challenging problem to build a robust KWS system that accurately wakes up anytime when the keyword is spoken but reliably suppresses most of the incoming negative audio.
Adversarial examples as a free resource have been widely applied to improve model robustness in different tasks by attacking model vulnerability \cite{kurakin2016adversarial} \cite{madry2017towards} \cite{pang2020boosting}.
They are crafted by adding imperceptible perturbations to mislead a well-trained neural network model \cite{carlini2018audio}.
Researchers have found that models trained with adversarial examples exhibit unexpected benefits, such as meaningful feature representations that align better with salient data characteristics \cite{freelunch} as well as enhanced robustness to corruptions concentrated in the high-frequency domain \cite{yin2019fourier}. To better leverage adversarial examples for training, Xie et al., have proposed to use an auxiliary batchnorm (BN) specifically for the adversary \cite{xie2020adversarial}. The statistics of the original and adversarial data distributions hence could be more accurately estimated for effective modeling.

In this work, we focus on how to effectively apply the generated adversarial examples to improve the robustness of a small-footprint KWS system. While \cite{wang2019adversarial} has explored to use adversarial examples as augmentation data to improve an attention-based KWS system, our interest concentrates on reducing the mismatch between the original training data and adversarial examples during training. Inspired by the effectiveness of the auxiliary BN approach in \cite{xie2020adversarial}, we propose datasource-aware disentangled adversarial training.
Specifically, we design a different auxiliary BN for each type of datasource in the training data and the corresponding adversarial attackers, such that the complementarity across different datasources could be fully exploited. The model architecture of our KWS system is built from the bottleneck residual block \cite{sandler2018mobilenetv2} that hinges on depth-wise separable convolutions \cite{sifre2014rigid}. We further extend the residual block by injecting a parameter-free simple attention model (SimAM) \cite{yang2021simam}. SimAM leverages a 3-dimensional attention map to refine the intermediate feature map in a CNN layer, so as to effectively increase model capacity with negligible computation cost.
The experimental results have shown the effectiveness of the proposed training approach with adversarial examples. On the internal dataset, we have improved false reject rate by $40.31\%$ at $1\%$ false accept rate, compared to the strongest baseline without using adversarial examples. On the Google Speech Commands V1 dataset, our best KWS system has achieved $98.06\%$ accuracy.


\section{Model Architecture}
\label{sec:mothod}

\subsection{Bottleneck Residual Block}
\label{sec:MN7-45}
Our KWS model adopts the bottleneck residual block in the MobileNetV2 architecture \cite{sandler2018mobilenetv2} which is tailored to efficient and effective modeling on mobile devices. The bottleneck residual block hinges on the depth-wise separable convolution and is supported by the linear bottleneck transformation and inverted residual connection.

\begin{figure}[h]
  \centering
  \includegraphics[width=0.9\linewidth]{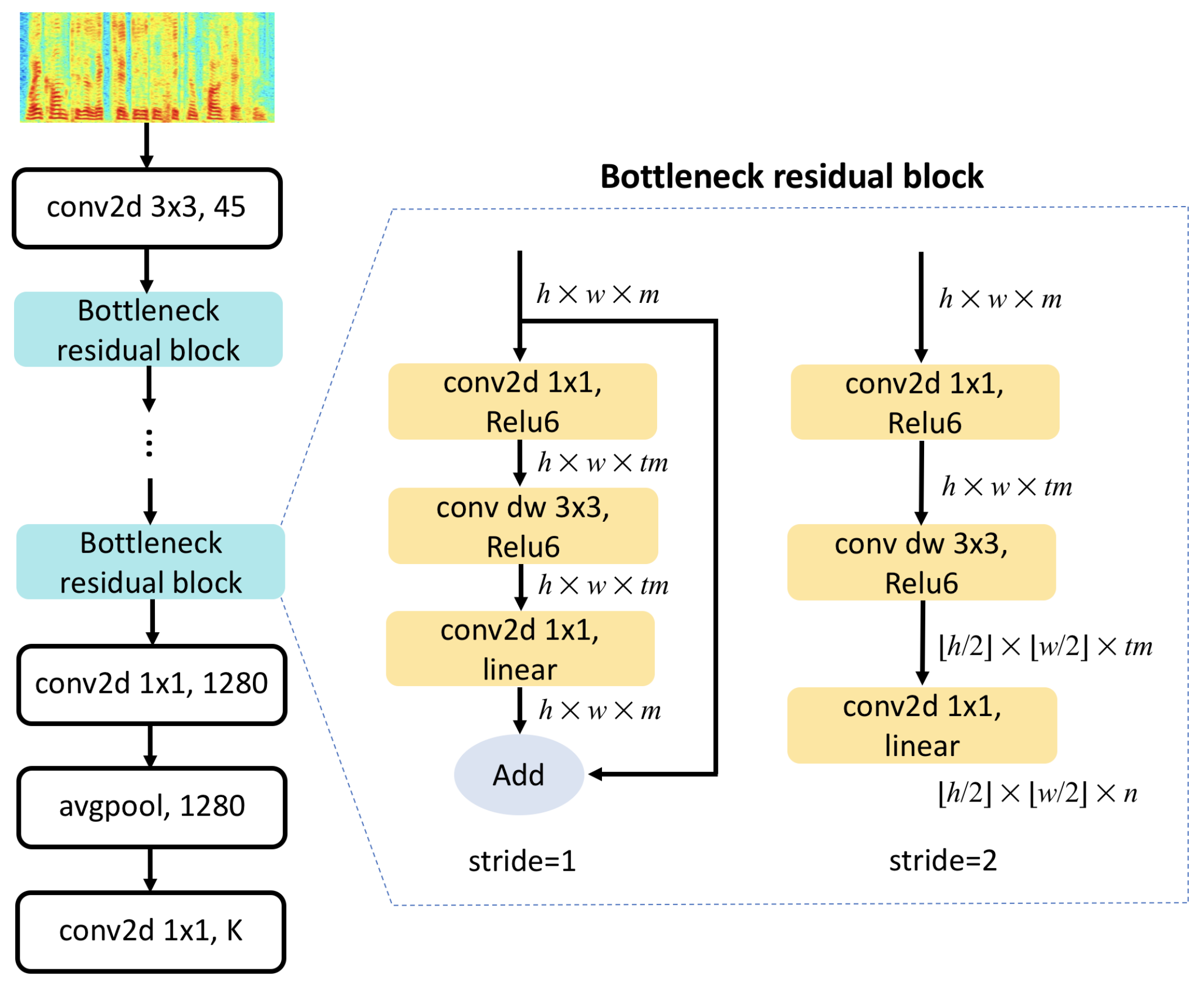}
  \caption{KWS model architecture (on the left) based on bottleneck residual layers (on the right) that take the input size $h \times w \times m$ and generate output size $\frac{h}{s} \times \frac{w}{s} \times n$.
  }
  \vspace{-1ex}
  \label{fig:diag}
\end{figure}

The structure of a bottleneck residual block is presented in Fig. \ref{fig:diag}. Each block is parameterized by its expansion factor $t$, the number of output channels $n$, and the stride $s$. Given an input of shape $h \times w \times m$, the initial pointwise convolution expands the input from $m$ channels to $tm$ channels followed by a $3 \times 3$ depth-wise convolution of stride $s$ as well as a linear bottleneck transformation.

The base model in this work is \texttt{MN7-45}, which is a variant of the MobileNetV2 architecture optimized for the KWS task.
The parameters of \texttt{MN7-45} are defined in Table \ref{tab:arc}. It consists of an initial standard convolution with $45$ filters followed by $7$ bottleneck residual blocks, global average pooling and a final output layer. Compared to MobileNetV2, \texttt{MN7-45} has fewer bottleneck residual blocks but with larger width in the initial layers, which we found to be more effective for the KWS task.

\begin{table}[h]
\fontsize{7.5}{10}\selectfont
\centering
\caption{KWS model architecture \texttt{MN7-45} with each layer described by the expansion factor $t$, the number of output channels $n$ and stride $s$.}
\label{tab:arc}
\begin{tabular}{c|c|c c c}
\toprule[2pt]
\multirow{2}{*}{\textbf{Layer}} & \multirow{2}{*}{\textbf{t}} & \multicolumn{3}{c}{\textbf{MN7-45}} 
\\ \cline{3-5} 
& & \textbf{n} & \textbf{s} & \textbf{Params} \\ \toprule[2pt]
conv2d 3x3 & - & 45 & 2 & 405 \\ 
bottleneck & 6 & 45 & 1 & 26.7K \\ 
bottleneck & 6 & 45 & 2 & 26.7K \\ 
bottleneck & 6 & 45 & 2 & 26.7K \\ 
bottleneck & 6 & 45 & 2 & 26.7K \\ 
bottleneck & 6 & 45 & 1 & 26.7K \\ 
bottleneck & 6 & 45 & 2 & 26.7K \\
bottleneck & 6 & 45 & 1 & 26.7K \\ 
conv2d 1x1 & - & 1280 & 1 & 57.6K \\ 
avgpool & - & 1280 & - & - \\ 
conv2d 1x1 & - & 2 & - & 2560 \\ \toprule[2pt]
\end{tabular}
\end{table}

\subsection{Simple Attention Module}
\label{sec:att}
Plug-and-play attention modules \cite{hu2018squeeze,woo2018cbam,yang2020gated} as an effective component can refine intermediate feature maps within a CNN block, so as to boost the model capacity.
The parameter-free simple attention module (SimAM) has shown the flexibility and effectiveness to improve the learning capacity of convolution networks with negligible computation cost \cite{yang2021simam}. It infers $3$-D attention weights for the feature map in a convolution layer by optimizing an energy function to capture the importance of each neuron. Specifically,
the minimal energy of a neuron $x$ in an input feature map $\vct{X}\in \mathbb{R}^{C\times H \times W}$ is expressed as:
\begin{align}
\begin{split}
e_x^\ast=\frac{4(\hat{\sigma}^2+\lambda)}{(x-\hat{u})^2+2\hat{\sigma}^2+2\lambda},
\label{eq:min_energy}
\end{split}
\end{align}
where $\hat{\mu}=\frac{1}{H \times W}\sum_{i=1}^{H\times W}x_i$, $\hat{\sigma}^2=\frac{1}{H \times W}\sum_{i=1}^{H\times W}(x_i-\hat{\mu})^2$, and $\lambda$ is a hyper parameter. The statistics $\mu$ and $\sigma$ are shared across all neurons within a channel, which hence significantly reduces computation cost.
As neuroscience study indicates an inverse correlation between the energy of $e_x^*$ and the importance of each neuron $x$ \cite{webb2005early}, the refinement of a feature map can be formulated as,
\begin{align}
\begin{split}
\hat{\vct{x}}=\sigma(\frac{1}{\mathbf{E}})\otimes\vct{x},
\label{eq:energy_dot_prod}
\end{split}
\end{align}
where $\mathbf{E}$ groups all energy values of $e^\ast_x$, and $\sigma(\cdot)$ denotes the sigmoid function. In this work, we plug a SimAM after the depthwise-wise convolution in each residual block of the base model \texttt{MN7-45}.


\section{Disentangled Adversarial Training}
\label{sec:DT}
\subsection{Adversarial Examples}
Adversarial examples are generated by adding imperceptible but malicious perturbations to the original data, such that the well-trained neural network can be misled to make an incorrect prediction \cite{goodfellow2014explaining}. In this work, we use the multi-step attacker based on Projected Gradient Descent (PGD) for adversarial data generation \cite{madry2017towards}.
Given an input training sample $\vct{x} \in \mathbb{D}$ with the corresponding ground-truth label $y$, a strong adversary is generated in an iterative manner,
\begin{align}
\begin{split}
\vct{x}_{t+1}^{adv}=\Pi_{\vct{x} + \mathbb{S}}(\vct{x}_t^{adv}+\mathop{\epsilon} \text{sgn}( \nabla_{\vct{x}}\mathrm{L}(\theta,\vct{x},y))),
\label{eq:pgd}
\end{split}
\end{align}
where $\Pi$ stands for a projection operator, $\mathbb{S}$ denotes the allowed perturbation space, $\epsilon$ is the step size, $\mathrm{L}(\cdot,\cdot,\cdot)$ is the loss function, and $\theta$ represents the model parameters. In our work, we use $8$ steps to generate an adversary. We treat the adversarial examples $\vct{x^{adv}}$ from Eq.~\ref{eq:pgd} as augmented data, and mix them with the original data for training, i.e.,
\begin{align}
\begin{split}
\mathop{\arg\min}_{\theta}[\mathbb{E}_{(\vct{x},y)\sim\mathbb{D}}(\mathrm{L}(\theta,\vct{x},y))+\mathrm{L}(\theta,\vct{x^{adv}},y)].
\label{eq:learning_obj}
\end{split}
\end{align}

\subsection{Disentangling via An Auxiliary BN}
\label{subsec:DAT}
Previous research work on adversarial attacks has found that training with adversarial examples could lead to label leaking, i.e., the neural network overfits to the specialized adversary distribution, resulting in degraded model performance \cite{kurakin2016adversarial} \cite{goodfellow2014explaining}.
To better leverage the regularization power of adversarial data, Xie et al., proposed disentangled training via an auxiliary batchnorm (BN)
to decouple the batch statistics between original and adversarial data in normalization layers during network training \cite{xie2020adversarial}, under the assumption that the adversarial examples and original data come from different underlying distributions.
At each training step, we maintain two BNs, i.e., one main BN and one auxiliary BN, respectively for the original mini-batch and the corresponding adversarial data, while the rest of network parameters are joint optimized for both data samples.
At the evaluation stage, we keep only the main BN by ignoring the auxiliary one.

\subsection{Fine-grained Disentangled Adversarial Training} 
Adversarial example generation can be easily generalized to a fine-grained version.
Instead of generating one adversarial example per training sample, we craft multiple adversaries at different perturbation levels ($\epsilon$ of Eq. \ref{eq:pgd} in a range of $[0.1,0.4]$) to capture a broader picture of training data statistics. Following basic disentangled learning in Section \ref{subsec:DAT},
we maintain one main BN for the original training data and a different auxiliary BN for the adversary at each perturbation level.

\subsection{Datasource-aware Disentangled Adversarial Training}
\begin{figure}[t]
  \centering
  \includegraphics[width=0.85\linewidth]{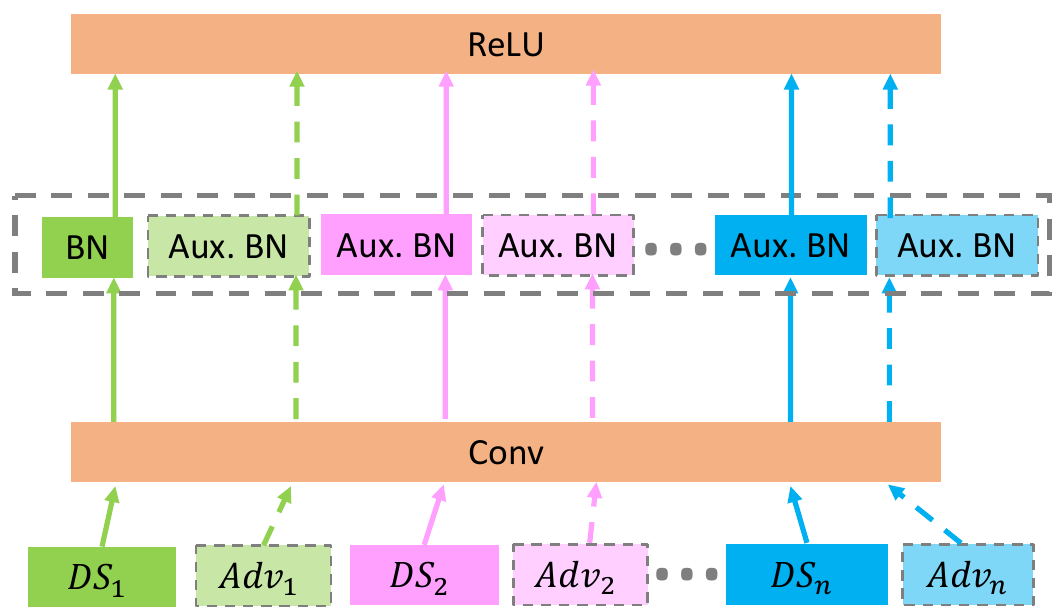}
  \caption{Information flow in the Data-source-aware fine-grained disentangled adversarial training architecture. (Boxes in solid colors denotes each data source, dashed boxes in corresponding faded colors are adversaries of each source)}
  \label{fig:DSA}
  \vspace{-0.2cm}
\end{figure}

We further extend the disentangling concept beyond the adversarial examples. The training data for KWS modeling usually has a multi-augmentation composition including clean audio, audio augmentation with background noise and speaking speed, as well as spectrum distortion with SpecAugment \cite{park2019specaugment}. Each augmentation type could be considered as one datasource. Our premise is that the distribution distinction exists not only between the adversary and original training data but also among the augmentation datasources within the original training data. The datasource-aware disentangled adversarial training hence aims to reduce all the possible mismatches. As illustrated in Fig.~\ref{fig:DSA}, we use one main BN for the clean data $DS_1$ and apply auxiliary BNs respectively for each augmentation datasource $DS_n$ ($n>1$) as well as for the corresponding adversarial examples $Adv_n$ ($n\geq 1$).
As described in Section \ref{subsec:DAT}, we keep the main BN at the evaluation stage by removing all the auxiliary BNs, assuming that the testing data is more likely to follow the clean data distribution rather than the artificial augmentation distribution.
The ablation study presented in Section \ref{sec:ablation} investigates the effectiveness of datasource-aware disentangled learning.


\section{Experiments}
\subsection{Data Description}
\label{subsec:datasets}
\paragraph*{Internal Dataset} The internal aggregated and de-identified multi-keyword dataset used in the experiments includes speech samples of four keywords, i.\,e., ``\emph{take a picture}'', ``\emph{volume down}'', ``\emph{volume up}'', and ``\emph{play music}'', collected through crowd-sourced workers. The training set contains $130$K positive samples with about $32$K in each keyword category and $100$K negative samples. The test set has $32$K positive samples with about $8$K for each keyword and $10$K negative samples. $6$ English accents are involved, including United States (US), Austraila (AU), Canada (CA), New Zealand (NZ), Britain (GB), and Latin America (LA). Training data includes the US accent only, and the other accents are used for testing.

\paragraph*{Google Speech Commands Dataset V1} The dataset consists of $1$s audio snippets recorded at sample rate $16$kHz in natural environments \cite{warden2018speech}, including $64,727$ samples of $30$ different words from $1,881$ speakers. The KWS model is trained to recognize $11$ classes: $10$ words ``\emph{up}'', ``\emph{down}'', ``\emph{left}'', ``\emph{right}'', ``\emph{yes}'', ``\emph{no}'', ``\emph{on}'', ``\emph{off}'', ``\emph{go}'', ``\emph{stop}'' as positive samples and the ``unknown'' category that includes the remaining words and background noise. We split the dataset into training, validation, and testing sets at the ratio of 8:1:1 by following the setting in \cite{warden2018speech} \cite{rybakov2020streaming}.

\subsection{Experimental Setup}
In the experiments, we consider two popular augmentation baselines: NoiseAugment and SpecAugment \cite{park2019specaugment}, in addition to the vanilla training using the clean data. The training data is distorted with various background noises (speech and music) at SNR sampled from [$0$dB, $+20$dB]. The testing data is also distorted at SNR of $10$dB and $20$dB in a similar manner but with different noise types.
We further use SpecAugment on top of the noise augmented training data. Training with adversarial examples (see Section \ref{sec:DT}) is applied to the $3$-datasource training data including clean, noisy, and SpecAugmented datasources. Specifically, we compare the performance of adversarial training without disentangled learning (\texttt{AT}) in Eq.~\ref{eq:learning_obj} \cite{wang2019adversarial}, disentangled adversarial training (\texttt{DAT}), fine-grained disentangled adversarial training (\texttt{FG\_DAT}), and datasource-aware disentangled adversarial training (\texttt{DA\_DAT}). We use \texttt{MN7-45} (see Section \ref{sec:MN7-45}) as the base model based on which we also investigate the effectiveness of SimAM for KWS modeling.

\begin{figure*}[t]
   \vspace{-4ex}
  \begin{minipage}[t]{0.49\textwidth}
  \centering
  \includegraphics[width=6.2cm]{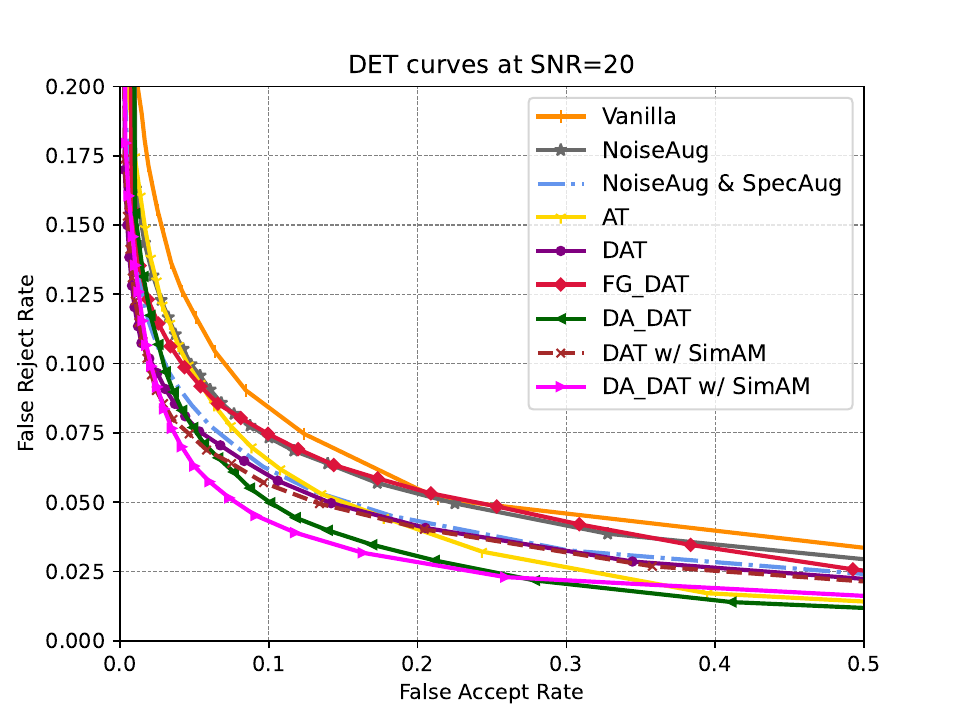}
  \end{minipage}
  \begin{minipage}[t]{0.49\textwidth}
  \centering
  \includegraphics[width=6.2cm]{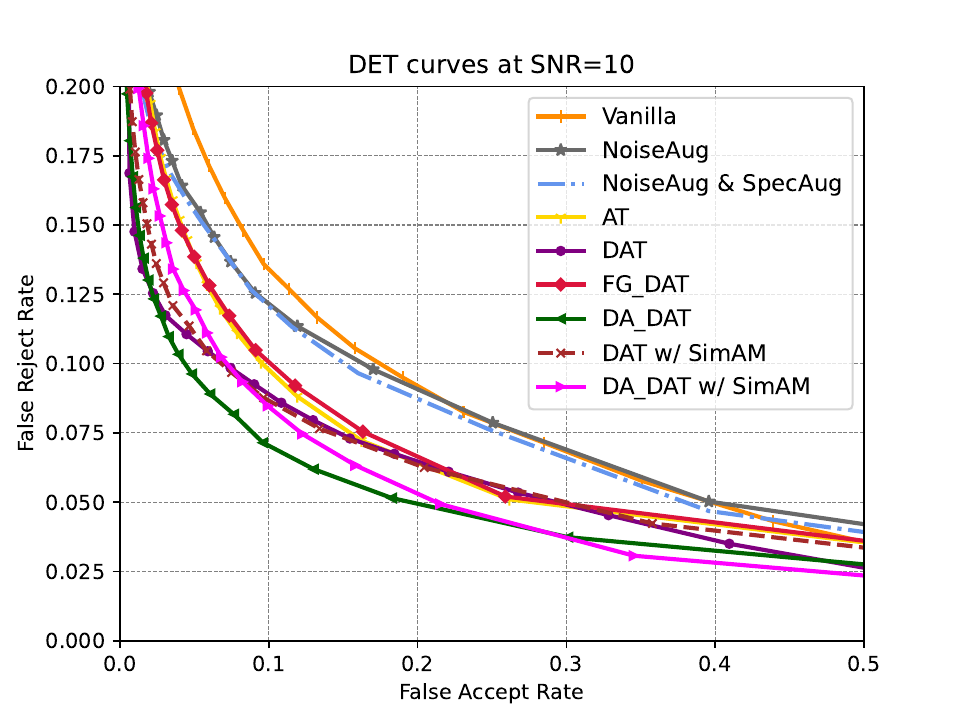}
  \end{minipage}
  \caption{DET curves of different approaches evaluated on testing data at SNR of $20$dB (left) and $10$dB (right)}
  \vspace{-3ex}
  \label{fig:roc}
\end{figure*}

We extract acoustic features using $40$-dim log Mel-filterbank energies that are computed over a $25$ms window every $10$ms.
In model training, we use an input window of $900$ms ($90$ frames) for the internal dataset and $1$s ($100$ frames) for Google speech commands dataset with the output of the corresponding keyword classes. The model inference is performed in a sliding window manner with a shift of $10$ frames in all experiments. We use the Adam optimizer with a learning rate of $0.005$ and cosine annealing learning rate decay.
We train the models for $15$ epochs.
We perform PGD attackers with perturbation strength $\epsilon$ in a range of $[0.1, 0.4]$. We find that $\epsilon = 0.2$ achieves the best performance on the internal dataset and $\epsilon = 0.1$ performs the best on the Google dataset.
The hyper-parameter $\lambda$ of SimAM in Eq.~\ref{eq:min_energy} is set to $0.0001$. False reject rate (FRR) and false accept rate (FAR) are used for internal data evaluation. We present experimental results by plotting detection error trade-off (DET) curves, where the $x$-axis and $y$-axis represent FAR and FRR, respectively.
Following the setup in the previous work \cite{warden2018speech} \cite{rybakov2020streaming}, we measure top-$1$ classification accuracy for the evaluation of the Google dataset.

\section{Results and Discussion}

\subsection{Results on Internal Data}
\label{sec:ablation}
Fig.~\ref{fig:roc} and Table \ref{table:adv} summarize the results of different approaches on testing data at SNR of $20$dB and $10$dB. We can observe that both NoiseAugment and its combination with SpecAugment consistently outperform the vanilla training on both distorted datasets. Adversarial examples with different training strategies could further improve the performance over the strongest baseline (NoiseAug+SpecAug). This performance boost is especially prominent on the highly noisy data. For example, \texttt{DA\_DAT} reduces FRR by $20.7\%$ at $1\%$ FAR on the $\textrm{SNR}=20\textrm{dB}$ data, and reduces FRR by $40.3\%$ on the $\textrm{SNR}=10\textrm{dB}$ data, compared to the baseline using NoiseAug+SpecAug. The results suggest the effectiveness of the adversarial examples for strengthening the generalizability of KWS models in the noisy environment.

\begin{table}[htbp]
\centering
\caption{False reject rate (FRR) of different approaches at $1\%$ false accept rate (FAR).}
\vspace{-0.1cm}
\label{table:adv}
\begin{adjustbox}{width=1.\columnwidth, center}
\begin{tabular}{c c c c c}
\toprule[2pt]
\multirow{2}{*}{\textbf{Methods}} & \multicolumn{2}{c}{$\textrm{SNR} = 20\textrm{dB}$} & \multicolumn{2}{c}{$\textrm{SNR} = 10\textrm{dB}$} \\

& \textbf{FRR} & \textbf{AUC} & \textbf{FRR} & \textbf{AUC} \\ 
\toprule[2pt]
Vanilla & $9.03\%$ & $0.956$ & $13.59\%$ & $0.939$ \\
NoiseAug & $7.33\%$ &$0.959$ & $12.55\%$ & $0.942$ \\ 
NoiseAug+SpecAug & $6.28\%$ & $0.968$ & $11.98\%$ & $0.944$ \\ 
\toprule[1pt]
AT & $6.20\%$ & $0.967$ & $10.04\%$ & $0.949$ \\ 
DAT & $5.77\%$ & $0.970$ & $8.59\%$ &$0.960$ \\ 
FG\_DAT & $7.47\%$ & $0.963$ & $10.48\%$ &$0.951$ \\ 
DA\_DAT & $4.98\%$ &$0.975$ & $\textbf{7.15}$\% & $\textbf{0.964}$ \\ 
\toprule[1pt]
DAT+SimAM & $5.70\%$ &$0.971$ & $8.77\%$ &$0.956$ \\
DA\_DAT+SimAM & $\textbf{4.52}\%$ &$\textbf{0.976}$ & $8.47\%$ &$0.961$ \\ 
\toprule[2pt]
\end{tabular}
\end{adjustbox}
\vspace{-0.8cm}
\end{table}

\paragraph*{Disentangled adversarial training}
Table \ref{table:adv} shows that \texttt{DAT} has exhibited substantial improvement over \texttt{AT}. We can see an FRR improvement of $7.23\%$ and $13.94\%$ respectively on the $\textrm{SNR}=20\textrm{dB}$ and $\textrm{SNR}=10\textrm{dB}$ data. This observation has corroborated that the data distribution mismatch does exist between the original data and the adversary, and disentangled learning with an auxiliary BN is beneficial for bridging such mismatch.
\vspace{-0.35cm}
\paragraph*{Datasource-aware disentangled adversarial training} 
We further compare the extensions of disentangled learning, i.e., \texttt{FG\_DAT} and \texttt{DA\_DAT}, against \texttt{DAT}. It is interesting to observe that \texttt{DA\_DAT} surpasses \texttt{DAT} on FRR by $13.69\%$ and $16.76\%$ respectively on the $\textrm{SNR}=20\textrm{dB}$ and $\textrm{SNR}=10\textrm{dB}$ data, while \texttt{FG\_DAT} performs worse than even \texttt{AT}. The performance gain from \texttt{DA\_DAT} indicates that the datasource-aware approach allows models to learn rich cross-datasource representations and hence increases model robustness against speech distortions.
In addition, the auxiliary BN method has demonstrated the effectiveness as a general adaptation technique across datasources. The inferior performance of \texttt{FG\_DAT} implies that one perturbation strength is often sufficient for effective attacks of one dataset while multiple perturbation strengths could bring confusion to modeling and degrades model performance.
\vspace{-0.35cm}
\paragraph*{Simple Attention Module} 
We update the structure of the base model \texttt{MN7-45} by adding a SimAM (see Sec. \ref{sec:att}) after the depth-wise convolution in each residual block. The updated model is trained using DAT and DA\_DAT which are the most effective training strategies. It has shown a marginal improvement with DA\_DAT on the $\textrm{SNR}=20\textrm{dB}$ data.

\subsection{Results on Google Speech Commands Data}
Table \ref{table:google} presents results on the Google Speech Commands data: the top-$1$ accuracy associated with the relative decrease in classification error rate (CER\_RD).
Similar to Section \ref{sec:ablation}, \texttt{FG\_DAT} obtains an inferior performance. Therefore we omit its results for simplicity.
We can observe that adversarial examples are helpful for boosting KWS performance. Specifically, the base model trained with \texttt{DA\_DAT} achieves an accuracy of $97.81\%$ with $18.28\%$ error rate reduction, compared to the baseline. Our best-performing system based on the SimAM module and trained with \texttt{DA\_DAT} achieves an accuracy of $98.06\%$.

\begin{table}[htbp]
\centering
\caption{Evaluation results on Google Speech Commands data.}
\vspace{-0.1cm}
\label{table:google}
\begin{tabular}{c c c}
\toprule[2pt]
\textbf{Methods} & $\textbf{Top-}\mathbf{1}$ \textbf{Acc}. & \textbf{CER\_RD}\\
\toprule[2pt]
NoiseAug+SpecAug & $97.32\%$ & -\\
\toprule[1pt]
AT & $97.64\%$ & $11.94\%$ $\color{green}\downarrow$ \\
DAT & $97.74\%$ & $15.67\%$ $\color{green}\downarrow$\\
DA\_DAT & $97.81\%$ & $18.28\%$ $\color{green}\downarrow$\\
DA\_DAT+SimAM & $\textbf{98.06\%}$ & $\textbf{27.61}\%$ $\color{green}\downarrow$\\
\toprule[2pt]
\end{tabular}
\vspace{-0.3cm}
\end{table}

\section{Conclusions}
In this paper, we explored how to effectively apply adversarial examples for KWS modeling. We proposed datasource-aware disentangled adversarial training through multiple auxiliary BNs. Experimental results on both internal dataset and Google Speech Commands dataset have demonstrated that adversarial examples as a free and infinite data resource could effectively boost KWS performance and the proposed training strategy exert the effectiveness to large extent.


\label{sec:conclusion}

\vfill\pagebreak

\newpage

\bibliographystyle{IEEEtran}
\bibliography{is2020_cross.bib}

\begin{thebibliography}{10}
\providecommand{\url}[1]{#1}
\csname url@samestyle\endcsname
\providecommand{\newblock}{\relax}
\providecommand{\bibinfo}[2]{#2}
\providecommand{\BIBentrySTDinterwordspacing}{\spaceskip=0pt\relax}
\providecommand{\BIBentryALTinterwordstretchfactor}{4}
\providecommand{\BIBentryALTinterwordspacing}{\spaceskip=\fontdimen2\font plus
\BIBentryALTinterwordstretchfactor\fontdimen3\font minus
  \fontdimen4\font\relax}
\providecommand{\BIBforeignlanguage}[2]{{%
\expandafter\ifx\csname l@#1\endcsname\relax
\typeout{** WARNING: IEEEtran.bst: No hyphenation pattern has been}%
\typeout{** loaded for the language `#1'. Using the pattern for}%
\typeout{** the default language instead.}%
\else
\language=\csname l@#1\endcsname
\fi
#2}}
\providecommand{\BIBdecl}{\relax}
\BIBdecl

\bibitem{miller2007rapid}
D.~R. Miller, M.~Kleber, C.-L. Kao, O.~Kimball, T.~Colthurst, S.~A. Lowe, R.~M.
  Schwartz, and H.~Gish, ``Rapid and accurate spoken term detection,'' in
  \emph{Eighth Annual Conference of the international speech communication
  association}, 2007.

\bibitem{103088}
J.~Wilpon, L.~Rabiner, C.-H. Lee, and E.~Goldman, ``Automatic recognition of
  keywords in unconstrained speech using hidden markov models,'' \emph{IEEE
  Transactions on Acoustics, Speech, and Signal Processing}, vol.~38, no.~11,
  pp. 1870--1878, 1990.

\bibitem{rohlicek1989continuous}
J.~R. Rohlicek, W.~Russell, S.~Roukos, and H.~Gish, ``Continuous hidden markov
  modeling for speaker-independent word spotting,'' in \emph{International
  Conference on Acoustics, Speech, and Signal Processing,}.\hskip 1em plus
  0.5em minus 0.4em\relax IEEE, 1989, pp. 627--630.

\bibitem{rose1990hidden}
R.~C. Rose and D.~B. Paul, ``A hidden markov model based keyword recognition
  system,'' in \emph{International Conference on Acoustics, Speech, and Signal
  Processing}.\hskip 1em plus 0.5em minus 0.4em\relax IEEE, 1990, pp. 129--132.

\bibitem{wu2018monophone}
M.~Wu, S.~Panchapagesan, M.~Sun, J.~Gu, R.~Thomas, S.~N.~P. Vitaladevuni,
  B.~Hoffmeister, and A.~Mandal, ``Monophone-based background modeling for
  two-stage on-device wake word detection,'' in \emph{2018 IEEE International
  Conference on Acoustics, Speech and Signal Processing (ICASSP)}.\hskip 1em
  plus 0.5em minus 0.4em\relax IEEE, 2018, pp. 5494--5498.

\bibitem{sun2017compressed}
M.~Sun, D.~Snyder, Y.~Gao, V.~K. Nagaraja, M.~Rodehorst, S.~Panchapagesan,
  N.~Strom, S.~Matsoukas, and S.~Vitaladevuni, ``Compressed time delay neural
  network for small-footprint keyword spotting.'' in \emph{Interspeech}, 2017,
  pp. 3607--3611.

\bibitem{chen2014small}
G.~Chen, C.~Parada, and G.~Heigold, ``Small-footprint keyword spotting using
  deep neural networks,'' in \emph{2014 IEEE International Conference on
  Acoustics, Speech and Signal Processing (ICASSP)}.\hskip 1em plus 0.5em minus
  0.4em\relax IEEE, 2014, pp. 4087--4091.

\bibitem{sainath2015convolutional}
T.~N. Sainath and C.~Parada, ``{Convolutional neural networks for
  small-footprint keyword spotting},'' in \emph{Proc. Interspeech 2015}, 2015,
  pp. 1478--1482.

\bibitem{tang2018deep}
R.~Tang and J.~Lin, ``Deep residual learning for small-footprint keyword
  spotting,'' in \emph{2018 IEEE International Conference on Acoustics, Speech
  and Signal Processing (ICASSP)}.\hskip 1em plus 0.5em minus 0.4em\relax IEEE,
  2018, pp. 5484--5488.

\bibitem{coucke2019efficient}
A.~Coucke, M.~Chlieh, T.~Gisselbrecht, D.~Leroy, M.~Poumeyrol, and T.~Lavril,
  ``Efficient keyword spotting using dilated convolutions and gating,'' in
  \emph{ICASSP 2019-2019 IEEE International Conference on Acoustics, Speech and
  Signal Processing (ICASSP)}.\hskip 1em plus 0.5em minus 0.4em\relax IEEE,
  2019, pp. 6351--6355.

\bibitem{kurakin2016adversarial}
A.~Kurakin, I.~Goodfellow, and S.~Bengio, ``Adversarial machine learning at
  scale,'' \emph{arXiv preprint arXiv:1611.01236}, 2016.

\bibitem{madry2017towards}
A.~Madry, A.~Makelov, L.~Schmidt, D.~Tsipras, and A.~Vladu, ``Towards deep
  learning models resistant to adversarial attacks,'' \emph{arXiv preprint
  arXiv:1706.06083}, 2017.

\bibitem{pang2020boosting}
T.~Pang, X.~Yang, Y.~Dong, K.~Xu, J.~Zhu, and H.~Su, ``Boosting adversarial
  training with hypersphere embedding,'' \emph{Advances in Neural Information
  Processing Systems}, vol.~33, pp. 7779--7792, 2020.

\bibitem{carlini2018audio}
N.~Carlini and D.~Wagner, ``Audio adversarial examples: Targeted attacks on
  speech-to-text,'' in \emph{2018 IEEE security and privacy workshops
  (SPW)}.\hskip 1em plus 0.5em minus 0.4em\relax IEEE, 2018, pp. 1--7.

\bibitem{freelunch}
D.~Tsipras, S.~Santurkar, L.~Engstrom, A.~Turner, and A.~Madry, ``There is no
  free lunch in adversarial robustness (but there are unexpected benefits),''
  \emph{arXiv:1805.12152}, 2018.

\bibitem{yin2019fourier}
D.~Yin, R.~Gontijo~Lopes, J.~Shlens, E.~D. Cubuk, and J.~Gilmer, ``A fourier
  perspective on model robustness in computer vision,'' \emph{Advances in
  Neural Information Processing Systems}, vol.~32, 2019.

\bibitem{xie2020adversarial}
C.~Xie, M.~Tan, B.~Gong, J.~Wang, A.~L. Yuille, and Q.~V. Le, ``Adversarial
  examples improve image recognition,'' in \emph{Proceedings of the IEEE/CVF
  Conference on Computer Vision and Pattern Recognition}, 2020, pp. 819--828.

\bibitem{wang2019adversarial}
X.~Wang, S.~Sun, C.~Shan, J.~Hou, L.~Xie, S.~Li, and X.~Lei, ``Adversarial
  examples for improving end-to-end attention-based small-footprint keyword
  spotting,'' in \emph{ICASSP 2019-2019 IEEE International Conference on
  Acoustics, Speech and Signal Processing (ICASSP)}.\hskip 1em plus 0.5em minus
  0.4em\relax IEEE, 2019, pp. 6366--6370.

\bibitem{sandler2018mobilenetv2}
M.~Sandler, A.~Howard, M.~Zhu, A.~Zhmoginov, and L.-C. Chen, ``Mobilenetv2:
  Inverted residuals and linear bottlenecks,'' in \emph{Proceedings of the IEEE
  conference on computer vision and pattern recognition}, 2018, pp. 4510--4520.

\bibitem{sifre2014rigid}
L.~Sifre and S.~Mallat, ``Rigid-motion scattering for texture classification,''
  \emph{arXiv preprint arXiv:1403.1687}, 2014.

\bibitem{yang2021simam}
L.~Yang, R.-Y. Zhang, L.~Li, and X.~Xie, ``Simam: A simple, parameter-free
  attention module for convolutional neural networks,'' in \emph{International
  conference on machine learning}.\hskip 1em plus 0.5em minus 0.4em\relax PMLR,
  2021, pp. 11\,863--11\,874.

\bibitem{hu2018squeeze}
J.~Hu, L.~Shen, and G.~Sun, ``Squeeze-and-excitation networks,'' in
  \emph{Proceedings of the IEEE conference on computer vision and pattern
  recognition}, 2018, pp. 7132--7141.

\bibitem{woo2018cbam}
S.~Woo, J.~Park, J.-Y. Lee, and I.~S. Kweon, ``Cbam: Convolutional block
  attention module,'' in \emph{Proceedings of the European conference on
  computer vision (ECCV)}, 2018, pp. 3--19.

\bibitem{yang2020gated}
Z.~Yang, L.~Zhu, Y.~Wu, and Y.~Yang, ``Gated channel transformation for visual
  recognition,'' in \emph{Proceedings of the IEEE/CVF conference on computer
  vision and pattern recognition}, 2020, pp. 11\,794--11\,803.

\bibitem{webb2005early}
B.~S. Webb, N.~T. Dhruv, S.~G. Solomon, C.~Tailby, and P.~Lennie, ``Early and
  late mechanisms of surround suppression in striate cortex of macaque,''
  \emph{Journal of Neuroscience}, vol.~25, no.~50, pp. 11\,666--11\,675, 2005.

\bibitem{goodfellow2014explaining}
I.~J. Goodfellow, J.~Shlens, and C.~Szegedy, ``Explaining and harnessing
  adversarial examples,'' \emph{arXiv preprint arXiv:1412.6572}, 2014.

\bibitem{park2019specaugment}
D.~S. Park, W.~Chan, Y.~Zhang, C.-C. Chiu, B.~Zoph, E.~D. Cubuk, and Q.~V. Le,
  ``Specaugment: A simple data augmentation method for automatic speech
  recognition,'' \emph{arXiv preprint arXiv:1904.08779}, 2019.

\bibitem{warden2018speech}
P.~Warden, ``Speech commands: A dataset for limited-vocabulary speech
  recognition,'' \emph{arXiv preprint arXiv:1804.03209}, 2018.

\bibitem{rybakov2020streaming}
O.~Rybakov, N.~Kononenko, N.~Subrahmanya, M.~Visontai, and S.~Laurenzo,
  ``Streaming keyword spotting on mobile devices,'' \emph{arXiv preprint
  arXiv:2005.06720}, 2020.

\end{thebibliography}

\end{document}